\documentclass{article}

\usepackage{spconf,amsmath,graphicx}
\usepackage{dblfloatfix}

\usepackage{newtxtext, newtxmath}
\usepackage{url}
\usepackage{xcolor}
\usepackage{arydshln}
\usepackage{enumerate}
\usepackage[mathcal]{euscript}
\usepackage{flushend}

\newtheorem{theorem}{Theorem}
\newtheorem{lemma}[theorem]{Lemma}

\newtheorem{corollary}{Corollary}

\newcommand{\vect}[1]{\mathbf{#1}}

\newcommand{\bs}[1]{\boldsymbol{#1}}
 
\graphicspath {{Figures/}}

\title{Human and Machine Type Communications can Coexist \\in Uplink Massive MIMO Systems}

\name{Kamil Senel, Emil~Bj\"{o}rnson and Erik~G.~Larsson\thanks{This work was supported by ELLIIT, by the Swedish Research Council (VR), project 2015-05573 and by Swedish Foundation for Strategic Research.}}
\address{Department of Electrical Engineering,\\ Link\"{o}ping University, Link\"{o}ping, Sweden\\
	Email: \{kamil.senel, emil.bjornson, erik.g.larsson\}@liu.se}

\begin{document}
\ninept
\maketitle
\begin{abstract}

Future cellular networks are expected to support new communication paradigms such as machine-type communication (MTC) services along with human-type communication (HTC) services. This requires base stations to serve a large number of devices in relatively short channel coherence intervals which renders allocation of orthogonal pilot sequence per-device approaches impractical. Furthermore, the stringent power constraints, place-and-play type connectivity and various data rate requirements of MTC devices make it impossible for the traditional cellular architecture to accommodate MTC and HTC services together. Massive multiple-input-multiple-output (MaMIMO) technology has the potential to allow the coexistence of HTC and MTC services, thanks to its inherent spatial multiplexing properties and low transmission power requirements. 
In this work, we investigate the performance of a single cell under a shared physical channel assumption for MTC and HTC services and propose a novel scheme for sharing the time-frequency resources. The analysis reveals that MaMIMO can significantly enhance the performance of such a setup and allow the inclusion of MTC services into the cellular networks without requiring additional resources.  
\end{abstract}

\vspace{-2mm}
\section{Introduction}
\label{sec:intro}
\vspace{-1mm}
One of the key technologies of $5$G future networks is the machine-type communications (MTC), which is projected to provide wireless connectivity to tens of billions of devices as a result of smart cities, factories, vehicles, and even common objects with sensing and communicating capabilities \cite{jeffrey2017towardsMTC}. A potential solution for accommodating the emerging traffic is utilizing the already existing infrastructure of cellular networks.~The standardization 
of techniques for MTC over cellular networks is already being considered \cite{3gpp2015mtcStandardizations}. However, the existing cellular network architectures, which are optimized to handle human-type communications (HTC), must be modified in order to handle HTC along with MTC, which requires consideration of a diverse communication characteristics \cite{metis2015report}. 

There are crucial problems to solve before the successful integration of MTC services into the existing cellular networks.~In particular, $5$G networks will have to support a large number of devices with low-complexity constraints and provide various data rates ranging up to multiple gigabits per second with reliability for services with stringent latency constraints such as health-care, security, and automotive applications \cite{metis2014mtc}. 

The performance of cellular networks in a setup where HTC and MTC services coexist has been considered in \cite{jeffrey2017towardsMTC,bontu2014IoT}.~WiFi-based networks constitute a competitive option to cellular networks and the integration of MTC services into the existing WiFi-based networks has been investigated \cite{sutton2017coexistWiFi}. A potential alternative is to utilize  multihop short-range transmission
technologies \cite{centenaro2016longRangeComm}. However, initial experimentations reveal the limitation of short-range technologies for MTC applications and emphasized the requirement of a plug-and-play type of connectivity without centralized planning which can be satisfied by long-range technologies  \cite{biral2015M2M}.    

A key technology of $5$G future cellular networks is MaMIMO in which the BSs are equipped with a large number of antennas, which gives them the ability to spatially multiplex multiple users \cite{redBook}. The advantages of MaMIMO technology have been shown to enhance the performance of cellular networks in terms of spectral efficiency and device detection in MTC setups \cite{jiang2016IotMIMO,elisabeth2017randomPilot,senel2017ampMTC}. However, to the best of authors' knowledge, this is the first work which considers the coexistence of HTC and MTC devices under a MaMIMO setup and analyze their joint spectral efficiency.

In this work, we investigate the performance, in terms of spectral efficiency, of a MaMIMO network that concurrently serves devices that utilize HTC and MTC. Different schemes for allocating time-frequency resources between MTC and HTC devices are investigated.~A novel resource allocation scheme is proposed and compared with the orthogonal and non-orthogonal resource allocation schemes. In particular, we answer the following questions:    
\begin{itemize}
	\item How will the existing cellular networks be affected by the dense MTC deployments?
	\item Can the challenges to accommodate MTC services over cellular networks be handled by the MaMIMO technology?
	\item Does the MaMIMO technology enable the use of non-orthogonal resources for MTC and HTC services thanks to its inherent utilization of spatial multiplexing? 
\end{itemize}

\vspace{-4mm}
\section{System Setup}\label{sec:SystemSetup}\vspace{-1mm}
We consider the uplink of a single-cell MaMIMO system where a BS with $M$ antennas is serving $K$ devices. Among these devices $K_m$ of them, referred to as \textit{machines}, require machine-type communication and the remaining $K_h = K - K_m$ devices, referred to as \textit{humans}, generate human-type traffic. Without loss of generality, assume that the indexes $k \in \mathcal{K}_h$, where $ \mathcal{K}_h =  \{1, \ldots, K_h\}$ are utilized for humans and $k \in \mathcal{K}_m = \{K_h + 1, \ldots, K\}$ denotes machines.

The time-frequency resources are divided into coherence intervals, such that each channel is constant and frequency-flat in each interval.~Each coherence interval (CI) has length $N$ (in samples) and a fraction of these samples are reserved for training whereas the remaining ones are utilized for uplink data transmission.

Non-line-of-sight communication is assumed and the small-scale fading is modeled as Rayleigh fading. Let $\vect{g}_k \in \mathbb{C}^{M \times 1}$ denote the channel between device $k$ and the BS, then
\begin{equation}
\vect{g}_k = \sqrt{\beta_k}\vect{h}_k, \forall k = 1,\ldots,K, 
\end{equation} 
where $\beta_k$ is the large-scale fading (same for all antennas) and $\vect{h}_k$ is the small-scale fading. Each element of $\vect{h}_k$ is modeled as i.i.d.$~\mathsf{CN}(0,1)$. The BS is assumed to know the large-scale fading coefficients. However, the small-scale fading coefficients are to be estimated in each CI. Moreover, $\vect{h}_k$ changes independently between CIs. 

\subsection{CI Allocation Schemes and Pilot Sequences} \label{sec:Schemes}
\begin{figure}[tb]
	\begin{center}
		\includegraphics[trim=0cm 0cm 0cm 0cm,clip=true, scale = .4]{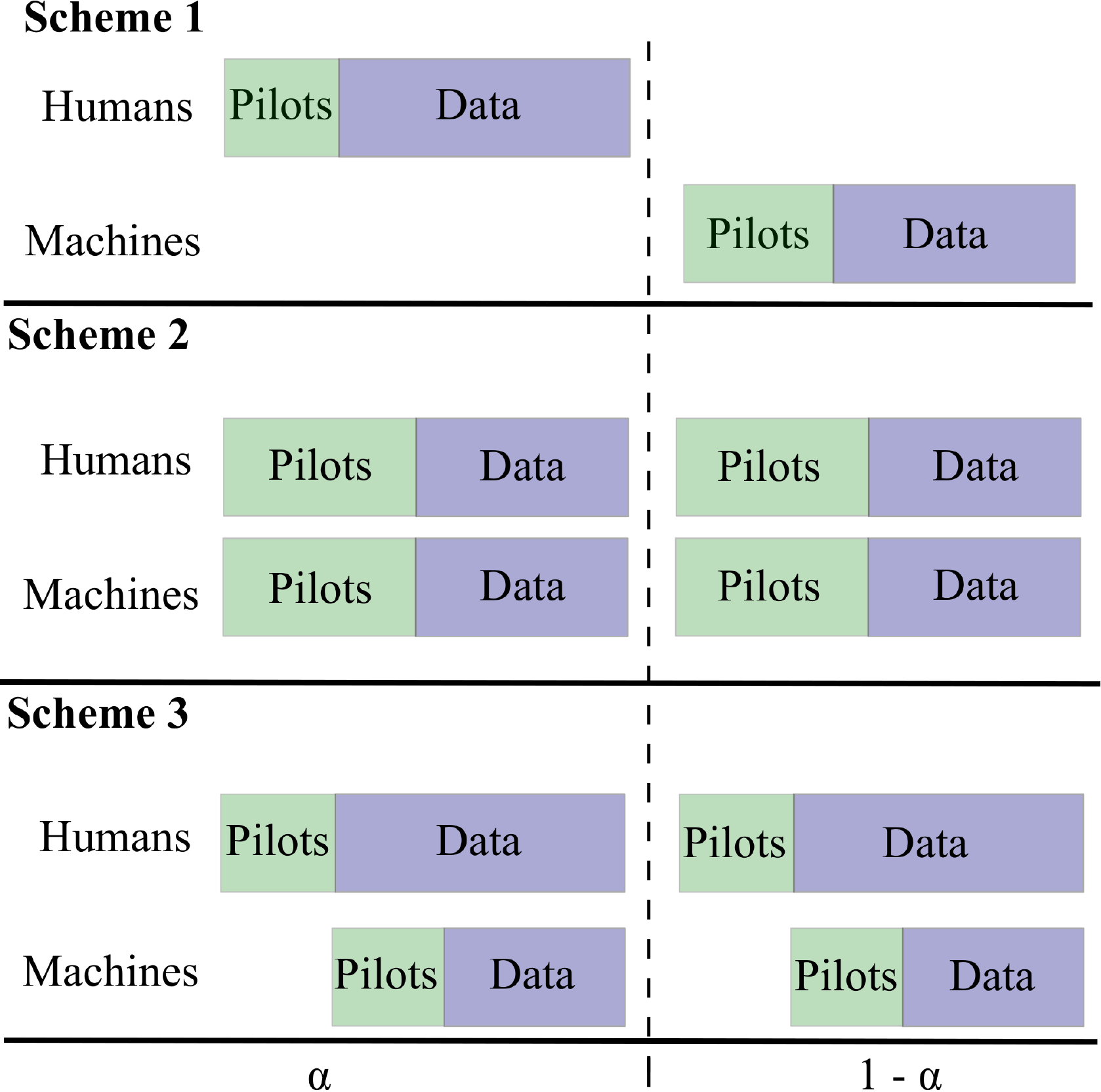}\vspace{-3mm}
		\caption{Coherence interval structure for training and data transmission for three different schemes. }
		\label{fig:CIs} \vspace{-8mm}
	\end{center}
\end{figure}

We consider three training and data transmission schemes.
\begin{itemize}
	\item \textbf{Scheme 1:} Humans and machines utilize different CIs.
	\item \textbf{Scheme 2:} All devices use the same training interval and data transmission interval in every CI. 
	\item \textbf{Scheme 3:} Machines are not allowed to transmit during the training period of humans, which reduces the human's pilot length. After the training of humans, machines transmit their pilot sequences followed by data transmission.   
\end{itemize}
Fig.~\ref{fig:CIs} illustrates the CI structures for the three schemes. Scheme $1$ is an orthogonal allocation scheme in the sense that it allocates different CIs to humans and machines. Scheme $2$ and $3$ are non-orthogonal schemes where both machines and humans utilize the same CIs. In Scheme $3$, we propose a novel approach by utilizing the training period of machines for data tranmission of humans. 

In MaMIMO setups, all $K$ devices concurrently transmit their pilot sequences during uplink training. The channels are estimated during uplink training and utilized during uplink data transmission. Traditionally, the pilot sequences are assumed to be orthogonal for users within a cell. These assumptions on pilot sequences are not realistic for massive MTC since it requires many pilots and cumbersome access procedures for pilot assignment \cite{elisabeth2017randomPilot}. Moreover, constraints on uplink power budget and excessive overhead signaling compel the use of non-orthogonal pilots for MTC \cite{jeffrey2017towardsMTC,elisabeth2017RandomAccessMag}.   

It is assumed that the number of machines is larger than the number of humans $K_m > K_h$ and it is not feasible to assign orthogonal pilots to machines. However, since the humans require higher data rates and are smaller in numbers compared to machines, we assume it is possible to allocate orthogonal pilot sequences to the $K_h$ humans. Let $\sqrt{N_p}\bs{\varphi}_k \in \mathbb{C}^{N_p \times 1}$ denote the $N_p$-length pilot signal for the $k$th device with $\|\bs{\varphi}_k\|^2 = 1$ and assume $k \in \{1, \ldots, K_h\}$, then
\begin{equation}\label{eq:PilotsHuman}
\bs{\varphi}_k^H\bs{\varphi}_i = 0,~~~ \forall k \in \{1, \ldots, K_h\}, \forall i \in \{1, \ldots, K\}. 
\end{equation}
For the MTC applications such as surveillance, machines also require high data-rates \cite{ratasuk2012mtc}. In these cases the machines with high data rate requirements may be treated as humans.

The pilot sequence length of humans, $N_p^h$, must be equal to or greater than the number of humans $K_h$ in order to allocate orthogonal pilots. On the other hand, 
machines are assumed to share $N_p^m$ orthogonal pilots. Each device picks one of the $N_p^m$ pilot sequences randomly in each coherence interval. 
Let $\bs{\varphi}_k$ be a pilot sequence assigned to a machine, then  
\begin{flalign}\label{eq:PilotMachine}
\bs{\varphi}_k^H\bs{\varphi}_i = \begin{cases}
0, &\text{if}~~i \in \mathcal{K}_h, \\
0, & \text{with Pr.~}1-\frac{1}{N_p^m},~~ \text{if}~~i \in \mathcal{K}_m, \\
1, & \text{with Pr.~}\frac{1}{N_p^m},~~\,~~~~~~ \text{if}~~i \in \mathcal{K}_m. 
\end{cases}
\end{flalign}
Although this pilot allocation avoids the need of a coordination process between the BS and the active devices in each CI, it introduces pilot contamination between machines.~Alternatively, orthogonal pilots can be assigned to active machines in each CI. However, this requires resources such as additional symbols and transmit power for coordination which are not desirable for MTC with latency and power constraints. 

\vspace{-2mm}
\section{Achievable Rate Analysis}\label{sec:SpectralAnalysis}
\vspace{-1mm}
In this section, the achievable rate of the three schemes illustrated in Fig.~\ref{fig:CIs} is investigated. Each scheme has an uplink training phase followed by data transmission.

\subsection{Analysis of Scheme 1}
In Scheme $1$ (SC-$1$), humans and machines utilize different CIs and therefore there is no interference between them. 
Let $\mathcal{K}$ be the set of active devices in a given CI. During the uplink training, active devices $k \in \mathcal{K}$ (In SC-$1$, either humans or machines are active, i.e., $\mathcal{K} = \mathcal{K}_h$ or $\mathcal{K} = \mathcal{K}_m$) concurrently transmit their pilot sequences and the composite received signal at the BS is 
\begin{equation}\label{eq:recSig-1}
\vect{Y} = \sqrt{N_p}\sum_{k' \in \mathcal{K}} \sqrt{q_{k'}}\vect{g}_{k'} \bs{\varphi}_{k'}^H + \vect{Z}
\end{equation}
where $\vect{Z} \in \mathbb{C}^{M \times N_p}$ is the noise matrix with i.i.d.$~\mathsf{CN}(0,\sigma^2)$ elements and $q_{k'}$ denotes the transmission power of pilot symbols for user $k'$.

In order to estimate the channel of user $k$, the BS performs a de-spreading operation on the received signal:
\begin{eqnarray}\label{eq:despred-1}
\vect{y}_k = \vect{Y}\bs{\varphi}_k 
= \sqrt{N_p q_k} \vect{g}_k + \sqrt{N_p}\hspace{-0.2cm}\sum\limits_{k' \in \mathcal{K}\backslash\{k\}} \hspace{-0.2cm}\sqrt{q_k}\vect{g}_{k'} \bs{\varphi}_{k'}^H\bs{\varphi}_k  + \vect{z}'
\end{eqnarray} 
where $\vect{z}' = \vect{Z}\bs{\varphi}_k$ has i.i.d.$~\mathsf{CN}(0,\sigma^2)$ elements since $\|\bs{\varphi}_k\|^2 = 1$. Then, the BS utilizes the linear minimum mean-square error (LMMSE) estimator to acquire $\hat{\vect{g}}_k$, the estimate of the channel between the BS and user $k$, as follows:
\begin{equation}\label{eq:MMSEest-1}
\hat{\vect{g}}_k = \frac{\sqrt{N_pq_k}\beta_k}{N_p\sum\limits_{k' \in \mathcal{K}}q_{k'}\beta_{k'}|\bs{\varphi}_{k'}^H\bs{\varphi}_{k}|^2 + \sigma^2}\vect{y}_k, 
\end{equation}  
(which is the true MMSE estimate for Scheme-$1$ and Scheme-$2$). $\hat{\vect{g}}_k$ has $M$ i.i.d. elements and the mean-square of the $m$th component is 
\begin{align}\label{eq:MSgamma}
\gamma_{k} = \mathbb{E}_\vect{g}\left[\left|\left[\hat{\vect{g}}_k\right]_m\right|^2\right] = \frac{N_pq_k\beta_k^2}{N_p\sum\limits_{k' \in \mathcal{K}}q_{k'}\beta_{k'}|\bs{\varphi}_{k'}^H\bs{\varphi}_{k}|^2+ \sigma^2}.
\end{align} 
where $\mathbb{E}_\vect{g}$ denotes the expectation with respect to $\vect{g}$. 
Let $s_k$ denote the unit power symbol to be conveyed by device $k$. Then, device $k$ transmits $x_k = \sqrt{p_k} s_k$,
where $p_k$ is the data transmit power of device $k$. In order to detect the data symbols of the $k$th device, the BS employs the maximum ratio combining (MRC) with the combining vector
\begin{equation}\label{eq:precoder}
\hat{\vect{v}}_k = \frac{1}{\gamma_k\sqrt{M}}\hat{\vect{g}}_k
\end{equation}
to compute the inner product with the received signal, $\vect{y}$ as 
\begin{eqnarray}\label{eq:compositeSignalatBS}
\hat{\vect{v}}_k^H\vect{y} 
&=&  \hat{\vect{v}}_k^H\vect{g}_kx_k + \sum_{k' \in \mathcal{K}\backslash\{k\}} \hat{\vect{v}}_k^H\vect{g}_{k'}x_{k'} + \hat{\vect{v}}_k^H\vect{z}.
\end{eqnarray} 
Based on \eqref{eq:compositeSignalatBS}, the achievable rate of device $k$ is given by
\begin{equation}\label{eq:SEofHumanSC-0}
R_k = \alpha_k\left( \frac{N_d}{N} \right)\log_2\left(1+ \Gamma_k\right)	
\end{equation}
where 
\begin{equation}\label{eq:CIdivisionSC-0}
N_d = \begin{cases}
N- N_p^h,  &\text{if}~~ k \in \mathcal{K}_h,\\
N - N_p^m, &\text{if}~~ k \in \mathcal{K}_m.
\end{cases}
\end{equation}
Here, $N_p^h$ and $N_p^m$ are the pilot lengths for humans and machines, respectively. $\alpha_k \in [0,1]$ represents the fraction of CIs assigned to humans/machines. The effective SINR term in \eqref{eq:SEofHumanSC-0} is \cite{emil2017SEofSuperimposed}
\begin{equation}\label{eq:SEofMachineSC-0}
\Gamma_k = \begin{cases}
\frac{Mp_k}{ \frac{1}{\gamma_{k}}\left( \sum\limits_{k' \in \mathcal{K}_h}p_{k'}\beta_{k'} + \sigma^2 \right) }\,, &\text{if}~~ k \in \mathcal{K}_h,\\
\frac{Mp_k}{\frac{1}{\bar{\gamma}_{k}}\left( \sum\limits_{k' \in \mathcal{K}_m}p_{k'}\beta_{k'} + \sigma^2 \right) + \frac{M}{N_p^m}\hspace{-.1cm} \sum\limits_{k' \in \mathcal{K}_m^k}\hspace{-0.2cm}\frac{p_{k'}q_{k'}\beta_{k'}^2}{q_k\beta_{k}^2} } \,, &\text{if}~~ k \in \mathcal{K}_m,
\end{cases}
\end{equation}
where $\mathcal{K}_m^k = \mathcal{K}_m \backslash\{ k \}$ and \begin{equation}\label{eq:SEofMachineSC-0gamma}
\bar{\gamma}_k = \mathbb{E}_{\bs{\varphi}}\left\lbrace\frac{1}{\gamma_{k}}\right\rbrace^{-1} = \frac{N_p^mq_k\beta_k^2}{ N_p^mq_{k}\beta_{k}+ \hspace{-.2cm}\sum\limits_{k' \in \mathcal{K}_m^k}\hspace{-.2cm}q_{k'}\beta_{k'} + \sigma^2}. 
\end{equation}

\subsection{Analysis of Scheme 2} \label{sec:Scheme-1SE} 
In Scheme $2$, each device uses $N_p$ symbols for training and $N-N_p$ symbols for data.~To find the corresponding rate of device $k$, we utilize the bounding techniques given in \cite{redBook} and state the following:
\begin{lemma}\label{lem:SEofSC-1}
	The achievable rate of device $k$ under Scheme $2$ is 
	\begin{equation}\label{eq:SEofHumanSC-1}
	R_k = \left( \frac{N - N_p}{N} \right)\log_2\left(1+ \Gamma_k\right)	
	\end{equation}
	where $\Gamma_k$ is the effective SINR for device $k$ and is defined by  
	\begin{equation}\label{eq:SEofMachineSC-1}
	\Gamma_k = \begin{cases}
	\frac{Mp_k}{ \frac{1}{\gamma_{k}}\left( \sum\limits_{k' \in \mathcal{K}}p_{k'}\beta_{k'} + \sigma^2 \right) }\,, &\text{if}~~ k \in \mathcal{K}_h,\\
	\frac{Mp_k}{\frac{1}{\bar{\gamma}_{k}}\left( \sum\limits_{k' \in \mathcal{K}}p_{k'}\beta_{k'} + \sigma^2 \right) + \frac{M}{N_p^m}\hspace{-.1cm} \sum\limits_{k' \in \mathcal{K}^k_m}\hspace{-0.25cm}\frac{p_{k'}q_{k'}\beta_{k'}^2}{q_k\beta_{k}^2} } \,, &\text{if}~~ k \in \mathcal{K}_m,
	\end{cases}
	\end{equation} 
	where 
	\begin{equation} 
	\bar{\gamma}_k = \mathbb{E}_{\bs{\varphi}}\left\lbrace\frac{1}{\gamma_{k}}\right\rbrace^{-1}\hspace{-.25cm} = \frac{N_pq_k\beta_k^2}{ N_pq_{k}\beta_{k}\hspace{-.05cm}+ \hspace{-.05cm}\frac{N_p}{N_p - K_h}\hspace{-.15cm}\sum\limits_{k' \in \mathcal{K}_m^k}\hspace{-.2cm}q_{k'}\beta_{k'} + \sigma^2}, \forall k \in \mathcal{K}_m. \nonumber 
	\end{equation}
\end{lemma}

The effective SINRs given by \eqref{eq:SEofMachineSC-1} reveals that as long as orthogonal pilots are assigned to humans, the integration of machines to the existing network does not create coherent interference, i.e., interference that scales with the number of antennas. Hence, as $M$ grows, the additional interference  originating from machines vanishes. However, this is not the case for machines as they suffer coherent interference due to the use of non-orthogonal pilots.  

\subsection{Analysis of Scheme 3} \label{sec:Scheme-2SE} 
In Scheme $3$, the machines are silent during the training of humans and send their pilot sequences while humans are transmitting data.
The LMMSE channel estimate for machines is  
\begin{equation} \label{eq:ChannelEstimateMachine-SC-2}
\hat{\vect{g}}_k = \frac{\sqrt{N_p^mq_k}\beta_k}{ N_p^m\hspace{-0.2cm}\sum\limits_{k' \in \mathcal{K}_m}\hspace{-0.2cm}q_{k'}\beta_{k'}|\bs{\varphi}_{k'}^H\bs{\varphi}_{k}|^2 + \hspace{-0.2cm} \sum\limits_{k' \in \mathcal{K}_h}\hspace{-0.2cm}p_{k'}\beta_{k'} + \sigma^2} \vect{y}_k, 
\end{equation}
which is not the MMSE estimator since 
\begin{equation}
\vect{y}_k = \sum\limits_{k' \in \mathcal{K}_h} \hspace{-0.2cm}\vect{g}_{k'}\vect{x}_{k'}^H\bs{\varphi}_k +    \sqrt{N_p^m}\hspace{-0.2cm}\sum\limits_{k' \in \mathcal{K}_m} \hspace{-0.2cm}\sqrt{q_{k'}}\vect{g}_{k'} \bs{\varphi}_{k'}^H\bs{\varphi}_k  + \vect{z}'
\end{equation}
 is not Gaussian. 
Notice that human's data symbols introduce coherent interference from humans to machines and deteriorate their channel estimation quality.  
\begin{lemma}\label{lem:SEofSC-2}
	The achievable rate of device $k$ under Scheme $3$ is 
	\begin{equation}\label{eq:SEofHumanSC-2}
	R_k = \left(\frac{N_d}{N} \right)\log_2\left(1+ \Gamma_k\right)	
	\end{equation}
	where $N_d$ is the number of data symbols utilized by device $k$ 
	\begin{equation}
	N_d = \begin{cases}
	N-N_p^h,  &\text{if}~~ k \in \mathcal{K}_h,\\
	N- N_p^h - N_p^m, &\text{if}~~ k \in \mathcal{K}_m,
	\end{cases}
	\end{equation}
	and $\Gamma_k$ is given by  
\vspace{-1mm}
	\begin{equation}\label{eq:SEofMachineSC-2}
	\Gamma_k = \begin{cases}
	\frac{Mp_k}{\frac{1}{\gamma_{k}}\left( \sum\limits_{k' \in \mathcal{K}}\hspace{-0.2cm}p_{k'}\beta_{k'} + \sigma^2\right) }, &\hspace{-3mm}\text{if} \,k \in \mathcal{K}_h,\\
	\frac{Mp_k}{\frac{1}{\bar{\gamma}_k}\left(\sum\limits_{k' \in \mathcal{K}}\hspace{-0.2cm}p_{k'}\beta_{k'} + \sigma^2 \right) + \frac{M}{N_p^m}\left(\hspace{-0.02cm}\sum\limits_{k' \in \mathcal{K}_m^k}\hspace{-0.25cm}\frac{p_{k'}q_{k'}\beta_{k'}^2}{q_k\beta_k^2} + \hspace{-0.25cm}\sum\limits_{k' \in \mathcal{K}_h}\hspace{-0.2cm}\frac{p_{k'}^2\beta_{k'}^2}{q_k\beta_k^2}\right) } , &\hspace{-3mm}\text{if}\,k \in \mathcal{K}_m.
	\end{cases}
	\end{equation} 
\end{lemma}

In SC-$3$, humans start data transmission after training without waiting for the training of machines. Hence, the number of available data symbols for humans is higher in SC-$3$ which comes at a cost of causing coherent interference to the machines.
   
\subsection{Asymptotic Analysis}
\label{sec:Asymptotic}
In order to gain further insights, the asymptotic limits of the rate expressions as $M \rightarrow \infty$ are investigated in this section. 
Note that as $M \rightarrow \infty$, the rate of humans, $R_k \rightarrow \infty$, $\forall k \in \mathcal{K}_h$ in all of the schemes considered. This is to be expected as humans suffer no coherent interference thanks to the orthogonal pilots allocated for them. For machines, the asymptotic limits are summarized as follows.

\begin{corollary}\label{cor:Asymptotic}
	The achievable SINR for device $k \in \mathcal{K}_m$ as $M~\rightarrow \infty$ is given by 
		\begin{equation}\label{eq:SEofMachineAsymptotic}
	\Gamma_k = \begin{cases}
	~~ \frac{p_k}{\hspace{-.1cm} \sum\limits_{k' \in \mathcal{\bar{K}}_m}\hspace{-0.05cm}\frac{p_{k'}q_{k'}\beta_{k'}^2}{{N_p^m}q_k\beta_{k}^2}}, &\text{for SC-1 and SC-2}, \\
	\frac{p_k}{ \frac{1}{N_p^m}\left(\sum\limits_{k' \in \mathcal{K}_m^k}\hspace{-0.25cm}\frac{p_{k'}q_{k'}\beta_{k'}^2}{q_k\beta_k^2} + \hspace{-0.15cm}\sum\limits_{k' \in \mathcal{K}_h}\hspace{-0.2cm}\frac{p_{k'}^2\beta_{k'}^2}{q_k\beta_k^2}\right) } , &\text{for SC-3.}
	\end{cases}
	\end{equation}
\end{corollary}  
 
The proof follows from taking the limit in \eqref{eq:SEofMachineSC-0}, \eqref{eq:SEofMachineSC-1} and \eqref{eq:SEofMachineSC-2}. 
The asymptotic analysis shows that SC-$1$ and SC-$2$ are equivalent in terms of asymptotic SINR whereas in SC-$3$ machines suffer additional coherent interference originating from humans. In both cases, the effective SINR increases with the pilot length of machines.    

\vspace{-2mm}
\section{Numerical Results} 

In this section, numerical results are presented which validate the analytical results in Section \ref{sec:SpectralAnalysis} along with the simulations that compare the schemes introduced in Section \ref{sec:SystemSetup}. The simulation setup consists of a cell with radius $250\,$m serving $K_h = 5$ humans and $K_m = 15$ machines which are uniformly distributed with a guard interval of $20\,$m. The path loss at distance $d\,$(km) is given by $130 + 37.6 \log_{10}(d)$ in dB. The power spectral density of the noise is $-174\,$dBm/Hz and the maximum transmit power for each device is $30\,$dBm.  

\begin{figure}[tb]
	\begin{center}
		\includegraphics[trim=.5cm .15cm 0cm .6cm,clip=true, width = 9cm,height = 6.1cm]{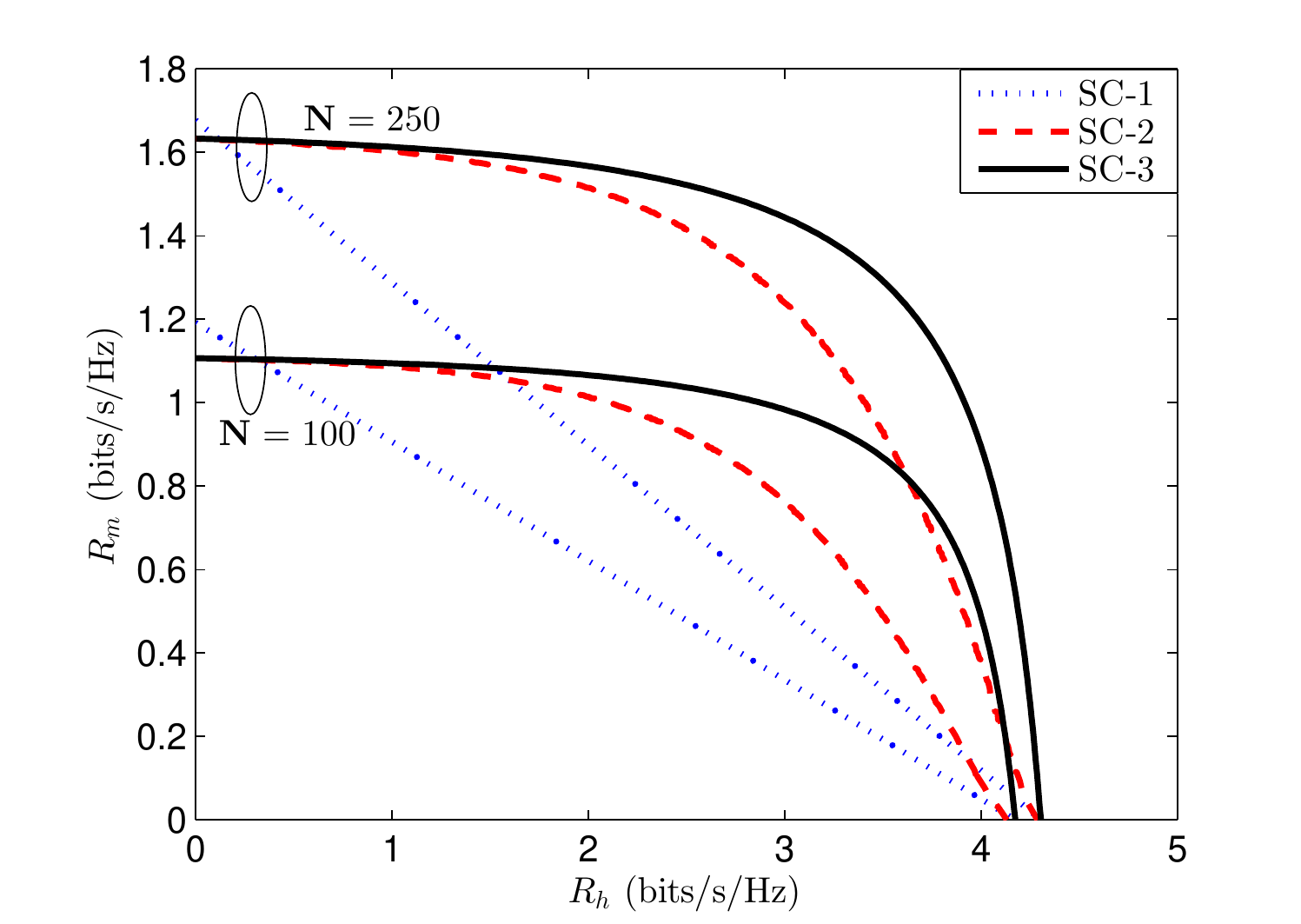}
	\vspace{-6mm}	\caption{Rate regions for max-min rates obtained via different schemes for $20$ devices with $M = 100$.}
		\label{fig:RateRegions} \vspace{-8mm}
	\end{center}
\end{figure}

Fig.~\ref{fig:RateRegions} depicts the rate regions for the schemes described in Section~\ref{sec:SpectralAnalysis}. Here, $R_h$ and $R_m$ denotes the max-min rate for humans and machines respectively. The rate curves are obtained by maximizing the minimum rate with respect to the transmit powers and the machine pilot length $N_p^m$. The pilot length of humans is fixed at $N_p^h = K_h$. The rate regions are obtained for two different CI lengths, which illustrates that Scheme-$1$ performs the best when there is only a single type of active device, i.e., either only humans or machines, in a given CI. However, for the cases where machines and humans coexist, allowing transmission from both results in a higher ergodic achievable rate as illustrated by  
Scheme-$2$ and Scheme-$3$. Furthermore, the numerical analysis shows that the novel Scheme-$3$ outperforms the other schemes under a heterogeneous setup.
    
\begin{figure}[tb]
	\begin{center}
		\includegraphics[trim=.5cm .05cm 0cm 0.6cm,clip=true, width = 9cm,height = 6.2cm]{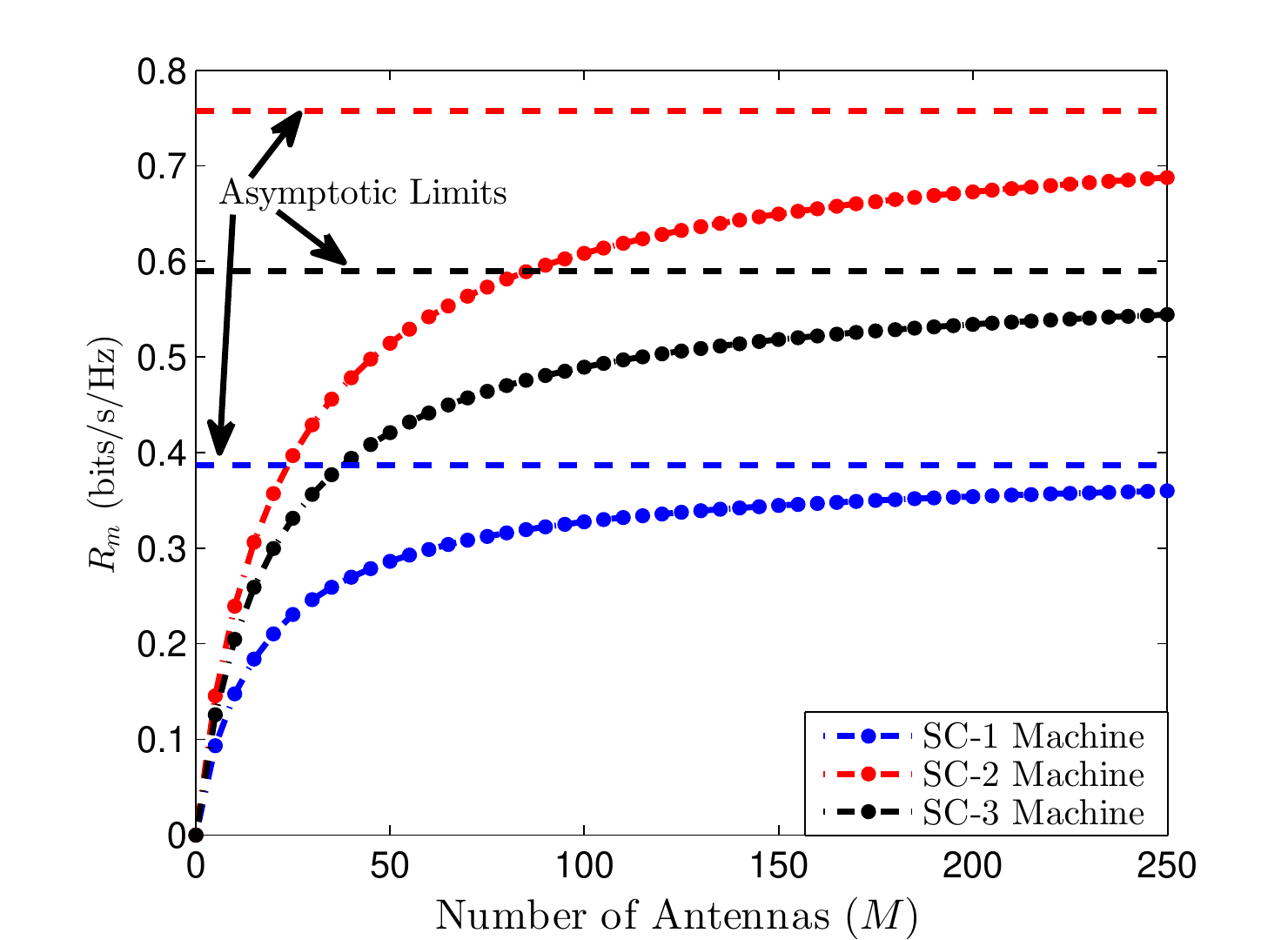}
	\vspace{-6mm}	\caption{Achievable ergodic rate and asymptotic limits for machines with each Scheme respect to number of antennas.}
		\label{fig:AsymptoticAnalysis} \vspace{-6mm}
	\end{center}
\end{figure}

The asymptotic limits provided in Corollary \ref{cor:Asymptotic} and the ergodic achievable rate with respect to the number of BS antennas are depicted in Fig.~\ref{fig:AsymptoticAnalysis}. Note that in each of the schemes $R_h \rightarrow \infty$ as $M \rightarrow \infty$ and therefore only the ergodic achievable rates of machines are included in the simulations which illustrates that even for relatively low number of antennas the difference between the asymptotic and achieved rate is small. This deviates from the conventional asymptotic results in MaMIMO, which require a very large number of antennas to reach asymptotic limits \cite{emil2017SEofSuperimposed}.   

\begin{figure}[tb]
	\begin{center}
		\includegraphics[trim=.5cm 0cm 0cm .5cm,clip=true, width = 9cm,height = 6.3cm]{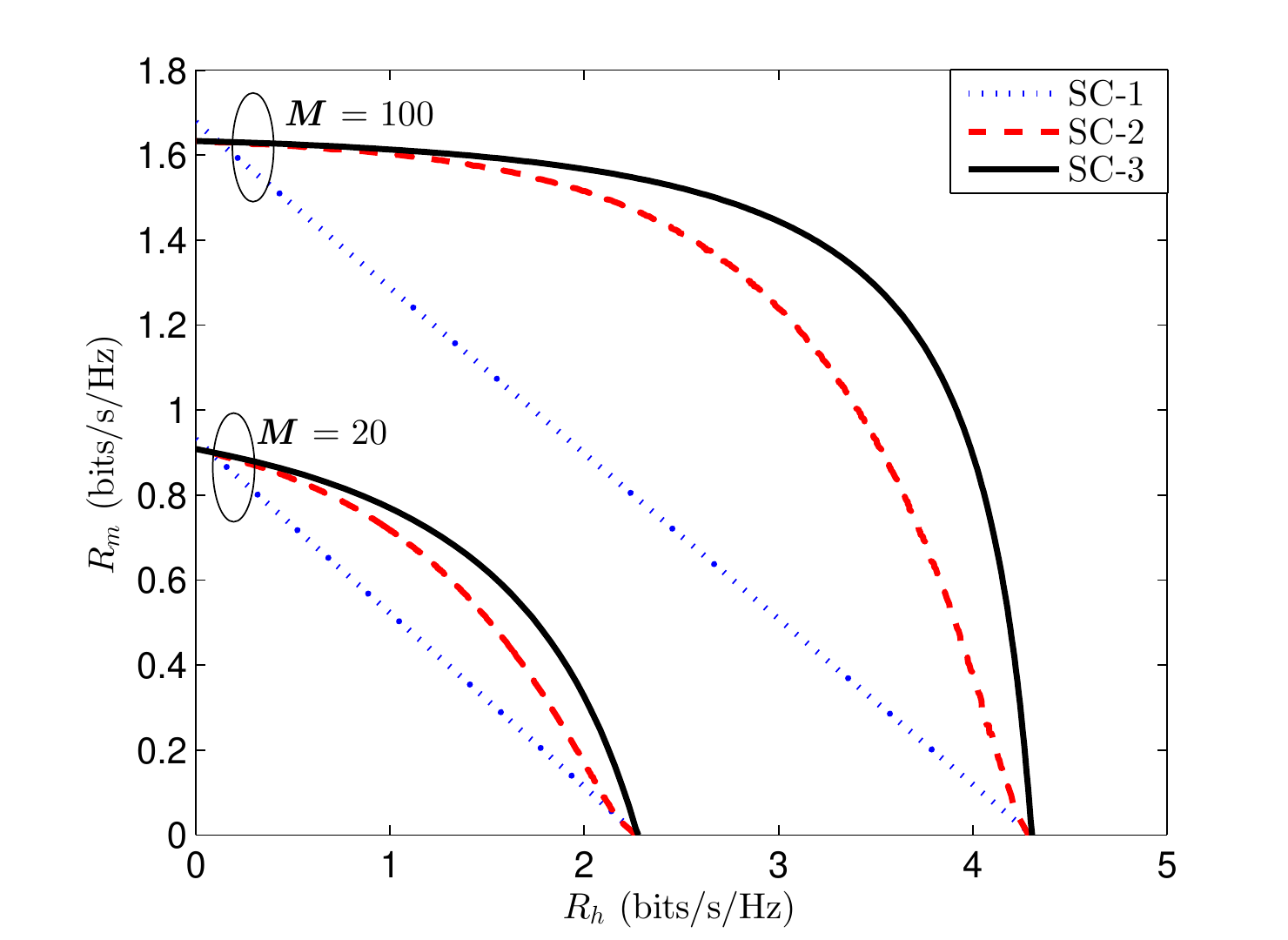} 
		\vspace{-8mm}\caption{Rate regions for max-min rates obtained via different schemes for $20$ devices with $N = 250$.}
		\label{fig:rateWrtM} \vspace{-9mm}
	\end{center}
\end{figure} 
 
The impact of the number of antennas is illustrated in Fig.~\ref{fig:rateWrtM}. As $M$ increases the non-orthogonal schemes (SC-$2$, SC-$3$) outperform orthogonal scheme (SC-$1$) due to two important reasons. First, the interference between humans and machines decreases with $M$, effectively converging to the SINR in the orthogonal scheme as $M \rightarrow \infty$. Also, the pre-log factor becomes dominant with increasing $M$ as the system starts to operate in bandwidth limited region at higher $M$ values contrary to lower $M$ values in which the system is in power limited region.  
  
\vspace{-2mm}
\section{Conclusion}

In this work, we consider the problem of accommodating both MTC and HTC in a MaMIMO system and proposed a novel resource allocation scheme along with the analysis of achievable rates for HTC and MTC. The analysis shows that MaMIMO is a key technology for enabling MTC in cellular networks thanks to the large array gain and spatial multiplexing.

\end{document}